%% file: sensitivity.tex
\documentclass[twocolumn,resetfootnote]{aastex701}
\usepackage{amsmath}
\usepackage{gensymb}

\begin{document}

\title{Pulsar Timing Array Sensitivity to Anisotropy: Empirical Sensitivity Curves, Scaling Relations, and the Multi-Resolution Pixel Basis}

\author[0009-0003-3700-446X]{Taha T. Moursy}
\affiliation{Department of Physics, Texas Tech University, Box 41051, Lubbock, TX 79409, USA}
\email{}
\author[0000-0002-8826-1285]{Nihan S. Pol}
\affiliation{Department of Physics, Texas Tech University, Box 41051, Lubbock, TX 79409, USA}
\email{}
\correspondingauthor{Taha T. Moursy}
\input{authors.tex}

\collaboration{all}{The NANOGrav collaboration}

\begin{abstract}

We quantify pulsar timing array (PTA) sensitivity to anisotropy in the gravitational wave background using the cross-correlation based Fisher information matrix in the pixel and spherical harmonic bases. We use a set of simulations to empirically determine scaling relations of a PTA's sensitivity to anisotropy with the number of pulsars $N_\mathrm{psr}$ in the array, the error $\delta t$ on the times of arrival, the frequency $f_\mathrm{GW}$ of the gravitational waves, and the angular scale $\Delta\Omega$ of the anisotropy. The sensitivity scales approximately as $N_\mathrm{psr}^{0.8}$, $\delta t^{-0.08}$, and $\Delta\Omega^{1.6}-\Delta\Omega^{2.1}$ (depending on the ranges of $\ell$ and $m$ under consideration). In addition, we use realistic simulations to project the NANOGrav PTA sensitivity to a 30-year baseline and quantify the growth in sensitivity at several timeslices. Except at the lowest frequencies, we find negligible effect on sensitivity through increasing the observation duration only. Finally, we introduce a multi-resolution pixel basis motivated by the large dependence of the sensitivity on sky location, and demonstrate the operation of the basis through a set of injections and recoveries.

\end{abstract}


\section{Introduction} \label{introduction}

Supermassive black holes are thought to reside at the center of many massive galaxies \citep{Richstone98}. When these galaxies merge, the black holes are brought closer and may become gravitationally bound as a supermassive black hole binary (SMBHB). 
These systems harden due to interactions with nearby stars and other matter \citep{smbhb_hardening_begelman}, and as the separation of these systems reaches $\sim$milli-parsec scales, they are visible via nano-Hertz frequency gravitational waves (GWs).
These low-frequency GWs are outside the band of ground-based GW detectors like the LIGO-Virgo-KAGRA \citep{LIGO,VIRGO,KAGRA} network but are detectable by pulsar timing array (PTA) experiments. 

PTAs are arrays of highly stable millisecond pulsars \citep{Backer82}, which emit beams of radio emission that can be observed using radio telescopes as the beams sweep across Earth with every rotation of the pulsar. Through precision timing of pulsar emission, a set of times of arrival (TOAs) can be generated for each pulsar \citep{Sazhin78, Detweiler79}. PTA collaborations compare these TOAs with predicted TOAs from a timing model which accounts for various deterministic and stochastic physical effects. For example, timing models include the pulsar's position and spin dynamics and the effects of the interstellar medium. The differences between the timing model and the observed TOAs are the timing residuals for a pulsar.

There can be several sources of nonzero residuals, such as random telescope-noise, clock calibration errors, poorly modeled solar wind, intrinsic pulsar noise, and, importantly, GWs. To distinguish a gravitational wave background (GWB) from other sources, PTA collaborations cross-correlate the residuals of the pulsars in the PTA and fit the correlations with the Hellings-Downs (HD; \citet{HD}) correlations model, which is the signature, ensemble-averaged set of correlations induced by an isotropic, stochastic GWB. The HD correlations, $\Gamma_{ab}^\text{HD}$, depend only on the angular separation $\xi_{ab}$ between pulsars $a$ and $b$,
\begin{equation}
\Gamma_{ab}^\mathrm{HD}=\frac{3}{2}(\frac{1-\cos\xi_{ab}}{2})\ln{\frac{1-\cos\xi_{ab}}{2}}-\frac{1-\cos\xi_{ab}}{8}+\frac{1}{2}+\frac{\delta_{ab}}{2},
\label{eq:HD_correlations}
\end{equation}
where $\delta_{ab}$ is the Kronecker delta.
These correlations distinguish the GWB from other sources which are either uncorrelated between pulsars, like pulse phase jitter, or have non-HD spatial correlations, like clock errors, which produce monopolar correlations in the residuals \citep{Hobbs12, Hobbs20}. 

In recent years, PTA collaborations have reported, with various levels of statistical significance, evidence for an HD-correlated GWB \citep{NG15gwb, EPTAgwb, PPTAgwb, CPTAgwb, MPTAgwb}.
The signal was consistent with being produced by a population of merging SMBHB systems \citep{NG15astro} in the local Universe, though more exotic, cosmological sources fit the data comparably well \citep{NG15np}.
These interpretations are not mutually exclusive, and the GWB could be the sum of an astrophysical GWB and a cosmological GWB. Nevertheless, there is interest in validating one or both interpretations of the GWB and placing constraints on the relevant astrophysical and cosmological models.

The two interpretations generally predict different power spectral densities (PSDs), so one method of distinguishing the origin is by making highly-constraining measurements and characterization of the GWB's PSD \citep{Kaiser22}. Another method is to use the spatial variation in the GWB. Cosmological GWBs may have some level of anisotropy due to, e.g., interference of GWs, but are generally statistically isotropic \citep{Caprini18}. On the other hand, astrophysical GWBs may have significant anisotropy due to loud, nearby SMBHBs which stand out against the background \citep{Mingarelli17}. Astrophysical GWBs may also have anisotropy which is correlated with the distribution of galaxies in the universe \citep{Mingarelli17}. Detecting anisotropy consistent with the SMBHB interpretation would be strong evidence in favor of that interpretation. 
A non-detection of anisotropy with PTAs would be in tension with the astrophysical interpretation and would favor the cosmological interpretation. Therefore, constraining anisotropy in the GWB is highly relevant for determining the GWB's origin.

The rest of this paper is organized as follows. In Section \ref{sec:methods}, we describe how to construct anisotropy sensitivity curves and how the simulations for this work were produced. In Section \ref{sec:results}, we present sensitivity curves made with simulated data as well as scaling relations of PTA sensitivity to anisotropy and realistic projections of the NANOGrav PTA sensitivity into the future. We discuss our results in Section \ref{sec:discussion} and end in Section \ref{sec:mesh} by introducing the new, multi-resolution pixel basis and presenting the results of injection-and-recovery simulations to validate the basis.

\section{Methods} \label{sec:methods}

\subsection{Anisotropy Methods and Sensitivity}

Anisotropy can be searched for in PTA data using cross-correlations to model the angular power distribution of the GWB \citep[e.g.][]{cornish_eigenmaps, Taylor13, Mingarelli13, Taylor20}.

The overlap reduction function (ORF) $\Gamma_{ab}$ given a GWB angular power density $\mathcal{P}(\hat\Omega)$ is \citep{Flanagan93},
\begin{equation}
    \Gamma_{ab}=\int_{S^2}\mathcal{R}_{ab}(\hat\Omega)\mathcal{P}(\hat\Omega)d\Omega,
    \label{eq:correlation}
\end{equation}
where $\mathcal{R}_{ab}(\hat\Omega)$ is the antenna response of pulsar pair $ab$ given by,
\begin{equation}
    \mathcal{R}_{ab}=\frac{3}{2}[\mathcal{F}_a^+\mathcal{F}_b^++\mathcal{F}_a^\times\mathcal{F}_b^\times],
    \label{eq:r_matrix}
\end{equation}
where $\mathcal{F}_a^A$ is the response of pulsar $a$ to $A$-polarized GWs,
\begin{equation}
    \mathcal{F}_a^A=\frac{\hat p_a^i\ \hat p_a^j}{1+\hat\Omega\cdot \hat p_a}e^A_{ij},
    \label{eq:f_matrix}
\end{equation}
where $\hat p_a$ is the position unit vector of pulsar $a$, $\hat\Omega$ is the GW propagation direction, and $e^A$ is the polarization basis tensor.

For an isotropic GWB, the power density $\mathcal{P}(\hat\Omega)$ is the same across the sky, $(4\pi \ \mathrm{sr})^{-1}$, and the cross-correlation for pulsar pair $ab$ is the sky average of $\mathcal{R}_{ab}(\hat\Omega)$, resulting in the HD correlations $\Gamma_{ab}^{\mathrm{HD}}$ in Eq.~\ref{eq:HD_correlations}.
An anisotropic GWB, on the other hand, has different power in different directions on the sky, and is thus represented by a more general distribution of the GWB power on the sky, $\mathcal{P}(\hat{\Omega})$.

Many bases exist to decompose the power, and in this work, we consider the radiometer pixel basis \citep{Ballmer06, Mitra08, Anholm09} and the spherical harmonic basis \citep{Mingarelli13, Taylor13}. The radiometer pixel basis uses delta function basis functions and a HEALPix\footnote{http://healpix.sf.net} \citep{HEALPix} tessellation of the sky to model the power. This basis assumes the power in each pixel is uncorrelated with the power in other pixels. The spherical harmonic basis expands the power $\mathcal{P}$ in real spherical harmonics up to some cutoff multipole. The estimator is then the spherical harmonic coefficients $c_{lm}$ in the expansion,
\begin{equation}
    \mathcal{P} = \sum_{l=0}^{l_\mathrm{max}}\sum_{m=-l}^{l} c_{lm}Y_{lm} \ .
\end{equation}
The real spherical harmonics $Y_{lm}$ are linear combinations of the spherical harmonics such that their imaginary components cancel.

Note that PTA anisotropy searches often use a power vector $\mathbf{P}$ with elements that are relative power densities in each pixel ($P_k = 4\pi \ \mathcal{P}(\hat\Omega_k)$), or $c_{lm}$ coefficients in the spherical harmonics basis, and define a response matrix $\mathbf{R}$ in such a way that the ORF is $\mathbf{\Gamma} = \mathbf{RP}$.

We use the per-frequency optimal statistic \citep[PFOS,][]{pfos} to compute pulsar pair cross-correlations $\rho_{ab}(f_k)$, their uncertainties $\sigma_{ab}(f_k)$, and the covariance matrix $\mathbf{C}(f_k)$ of the correlations, where $a$ and $b$ represent pulsars, and $f_k$ is a frequency in the Fourier basis used to model the GWB, $k/T_{\mathrm{span}}$ \citep{Lentati13}.

The cross-correlations are given by \citep{pfos}
\begin{equation}
    \rho_{ab}(f_k)=\frac{\boldsymbol{X}_a^T\tilde{\phi}(f_k)\boldsymbol{X}_b}{\mathrm{tr}[\boldsymbol{Z}_a\tilde{\phi}(f_k)\boldsymbol{Z}_b\Phi(f_k)]},
    \label{eq:pfos_rho}
\end{equation}
where $\mathbf{X}_a$ and $\mathbf{Z}_a$ are matrices which are constructed from the noise properties and models of pulsar $a$ \citep{Pol22,pfos}.

The cross-correlation uncertainties are given by
\begin{equation}
    \sigma^2_{ab}(f_k)=\frac{\boldsymbol{Z}_a\tilde{\phi}(f_k)\boldsymbol{Z}_b\tilde{\phi}(f_k)}{\mathrm{tr}[\boldsymbol{Z}_a\tilde{\phi}(f_k)\boldsymbol{Z}_b\Phi(f_k)]^2} \ .
\end{equation}
The pulsar-pair covariance matrix,
\begin{equation}
    \boldsymbol{C}(f_k)\equiv \langle \rho_{ab}(f_k)\rho_{cd}(f_k)\rangle - \langle \rho_{ab}(f_k)\rangle\langle \rho_{cd}(f_k)\rangle \ ,
\end{equation}
has elements derived and given in \citet{pfos}.
See \citet{pfos} for more details on the PFOS framework and the above quantities, including explicit expressions for $\mathbf{C}(f_k)$.

Given a cross-correlation vector $\boldsymbol{\rho}$ of length $N_{cc}$, the number of cross-correlations, the log-likelihood for $\boldsymbol{\rho}$, given a model power distribution of the GWB $P(\hat\Omega)$, is \citep{Romano17, AliHaimoud20, AliHaimoud21, Pol22}
\begin{equation}
\mathrm{log}\ \mathcal{L}(\boldsymbol{\rho}|\mathbf{P}) = -\frac{1}{2}(\boldsymbol{\rho} - \mathbf{RP})^T\mathbf{C}^{-1}(\boldsymbol{\rho}-\mathbf{RP}) + \mathrm{constant},
\end{equation}
where $\mathbf{P}$ is an $N_{\mathrm{pix}}$ ($N_{c_{lm}}$) vector containing the power decomposition of $P(\hat\Omega)$ in the radiometer (spherical harmonic) basis and $\mathbf{R}$ is a matrix related to the antenna response $\mathcal{R}_{ab}$ but with an extra normalization factor $N_{\mathrm{pix}}^{-1}$ for the pixel basis and convolved with the real spherical harmonics for the spherical harmonic basis. Explicitly,
\begin{equation}
    R_{ab}^k = \frac{1}{N_{\mathrm{pix}}}\mathcal{R}_{ab}(\hat\Omega_k)
\end{equation}
for the pixel basis, and
\begin{equation}
    R_{ab}^{lm}=\sum_{k=1}^{N_{\mathrm{pix}}}Y_{lm}(\hat\Omega_k)R_{ab}^k
\end{equation}
for the spherical harmonic basis.

Since these two bases express the power in linear combinations of the basis eigenfunctions, the maximum-likelihood solution for $\mathbf{P}$ can be found analytically by partial differentiation of log $\mathcal{L}$ with respect to $\mathbf{P}$ and solving for the root. The solution is
\begin{equation}
    \mathbf{P}=\mathbf{M}^{-1}\mathbf{X},
\end{equation}
where $\mathbf{M}$ is the Fisher information matrix given by $\mathbf{M}=\mathbf{R}^T \mathbf{C}^{-1}\mathbf{R}$, and $\mathbf{X}$ is the dirty map, given by $\mathbf{X}=\mathbf{R}^T \mathbf{C}^{-1}\boldsymbol{\rho}$. $\mathbf{X}$ here is unrelated to the pulsar-specific matrix $\mathbf{X}_a$ previously shown in Equation \ref{eq:pfos_rho}.
The covariance of the model parameters $P_i$ is the inverse of the Fisher information matrix, so the uncertainty on the regression parameters can be approximated by $\sqrt{\mathrm{diag}(\mathbf{M}^{-1})}$. For the radiometer pixel basis, this is a vector of $N_{\mathrm{pix}}$ elements giving the uncertainty in the recovered power in each pixel, and for the spherical harmonic basis, this is a vector of $N_{c_{lm}}$ elements giving the uncertainty in the $c_{lm}$ coefficients. To compute these matrices, we use the existing frameworks in \texttt{Defiant} \citep{pfos}, a software package for the PTA optimal statistic \citep{Anholm09,Demorest12,Chamberlin15}, and in \texttt{MAPS} \citep{Pol22}, a software package for PTA anisotropy searches and methods. Although we base our analyses on the previously-mentioned software, we accelerate the computations with \texttt{JAX} \citep{jax2018github}, among other optimizations, to achieve a speedup of $\sim$50,000 times (when run on an NVIDIA A100). The relevant code is available on \texttt{GitHub} as a fork\footnote{github.com/TTMoursy/defiant} of \texttt{Defiant}.

We use the uncertainties given by $\sqrt{1/\mathrm{diag}(\mathbf{M})}$ (i.e., neglecting covariances between basis functions) to quantify the sensitivity of a PTA to anisotropy in the GWB. In the radiometer basis, this gives a sky map of uncertainties. In the spherical harmonic basis, this gives a set of uncertainties at each angular scale corresponding to the $\ell$ value as well as each directional orientation corresponding to the $m$ value. In any basis, using the PFOS gives uncertainties as a function of frequency, and this information can be used to construct anisotropy sensitivity curves. 

\subsection{Simulations} \label{sec:simulations}

We wish to study how the uncertainties scale with properties of a PTA, namely the number of pulsars $N_{\mathrm{psr}}$, TOA error $\delta t$, and the GW frequency $f_{\mathrm{GW}}$ at which the search is performed. To this end, we use \texttt{libstempo}\footnote{github.com/vallis/libstempo} to generate simulated datasets based on the NANOGrav 15-year dataset.
We simulate two types of datasets.
The first set is used to determine scaling laws of the uncertainties with $N_{\mathrm{psr}}$, $\delta t$, $f_{\mathrm{GW}}$, and angular scale. We describe the details of the simulations below.
The second set is used to forecast the sensitivity of the NANOGrav PTA into the future by exploring the effect of increasing $T_{\mathrm{span}}$ on the sensitivity to anisotropy (we change $T_\mathrm{span}$ only and not $N_\mathrm{psr}$ or TOA uncertainty to isolate the effect of increasing the observation baseline).

For the first set of simulations, we aim to find scaling relations of PTA sensitivity to anisotropy with various PTA parameters. We again begin with the NANOGrav 15-year dataset. We simulate 100 datasets with different noise realizations for each combination of $N_{\mathrm{psr}}$ and TOA uncertainty in the sets 17, 34, 51, and 67 pulsars and 100 ns, 500 ns, and 1 \textmu s uncertainty, respectively, giving a total of 1200 datasets. 
For a given choice of TOA uncertainty, we set all TOAs in these datasets to have the same uncertainty.
In the datasets with less than 67 pulsars, we randomly select the subset of pulsars, and we use the same subset across levels of TOA uncertainty, but we use a different random selection for each realization of injected noise processes. 
For example, in the 17-pulsar case, we simulate 3 datasets with the same 17 pulsars but with 100 ns, 500 ns, and 1 \textmu s TOA uncertainties, and then another 3 datasets with a different random choice of 17 pulsars, and so on until there are 300 datasets in total which have 17 pulsars, and those 300 datasets in total contain 100 choices of 17 pulsars.
We use the same pulsar positions and timing models as the real dataset, but we extend the timing baseline of each pulsar to 20 years. Each pulsar thus has the same timing baseline but different start and end observation dates.
We use epoch-averaging to reduce the number of TOAs and thereby reduce the computational cost of analyzing the datasets \citep{a4c}. We inject intrinsic red noise (IRN) in each simulated pulsar with the values given in Table \ref{table:injections} of Appendix \ref{appendix:injected_noise}, which were measured from the real NANOGrav 15 yr dataset \citep{NG15timing}. We inject an isotropic GWB with an amplitude of $A_{1\mathrm{yr}} = 2.4\times10^{-15}$ and spectral index $\gamma = 13/3$, consistent with the reported values in the NANOGrav 15-year GWB search \citep{NG15gwb}.

For the second set of simulations, we generate 50 datasets with different realizations of measurement noise, IRN, and the GWB. 
To make each dataset, we begin with the NANOGrav 15-year dataset and extend it to 30 years of observation. The pulsars in this set of simulations do not have the same baselines as one another. We use the same pulsars and timing models again as in the real dataset, and we inject the same IRN and GWB values mentioned in the previous paragraph. The injected measurement noise is realistic in these datasets however, in contrast to the scaling law simulations, and we use the method of \citet{a4c} and the measured noise from the real dataset for these injections.

To analyze the datasets in each of the two sets previously described, we first perform common uncorrelated red-noise (CURN) broken power law analyses to determine the number of Fourier modes which have support for a common process. We use \texttt{ENTERPRISE} \citep{enterprise}, \texttt{enterprise\_extensions} \citep{enterprise_extensions}, and \texttt{PTMCMCSampler} \citep{ptmcmcsampler} for these analyses. We then perform CURN power law analyses using the previously determined number of Fourier modes to model the GWB. We again use the same software for these analyses. We use the resulting posteriors for IRN and the GWB spectral parameters to compute the noise-marginalized \citep{Vigeland18}, pair-covariant PFOS \citep{Gersbach25} and pulsar pair cross-correlation information. We convert the PFOS values $S_k$ to characteristic strain with $h_c(f_k) = \sqrt{12\pi^2f^3S_kT_{\mathrm{span}}}$ \citep{Rosado15}, where $T_{\mathrm{span}}$ is the total observation duration of the simulated PTA. We use the cross-correlation uncertainties to compute the Fisher matrices for the two bases being considered in this work and as described in Section \ref{sec:methods}. 

To obtain the uncertainty in $h_c(f_k)$ in each pixel of the radiometer basis, we multiply $h_c(f_k)$ by the relative uncertainty in the pixels, given by the square root of the diagonal elements of $\mathbf{M}^{-1}$. At this point, we have uncertainties on $h_c$ as a function of both frequency and sky location, $\sigma_{h_c}(f_k, \hat{\Omega})$. We produce sensitivity curves by selecting a sky location and tracking the uncertainty in $h_c$ at that location across frequencies. We can also similarly produce sky-averaged sensitivity curves by taking the sky-averaged uncertainty at each frequency. To compare our results with prior work on isotropic GWB sensitivity curves, we use the formalism implemented in \texttt{hasasia} \citep{hasasia_software} to calculate sensitivity curves for our simulated datasets. Specifically, we compute sensitivity curves for an isotropic stochastic GWB and compare them to map-averaged anisotropy sensitivity curves neglecting pair covariance.

For the spherical harmonic basis, the regression coefficients are the $c_{lm}$ values, so the square root of the diagonal elements of $\mathbf{M}^{-1}$ give the uncertainties on the spherical harmonic coefficients rather than the power in the pixels as in the radiometer pixel basis. We compute the Fisher matrices similarly to the process for the radiometer pixel basis and then analyze the scaling of the $c_{lm}$ uncertainties with frequency, number of pulsars, RMS TOA error, and the angular scale $\ell$.

\section{Results} \label{sec:results}

\subsection{Sensitivity Curves}\label{sensitivity_curves}

\begin{figure*}
    \includegraphics[width=\textwidth]{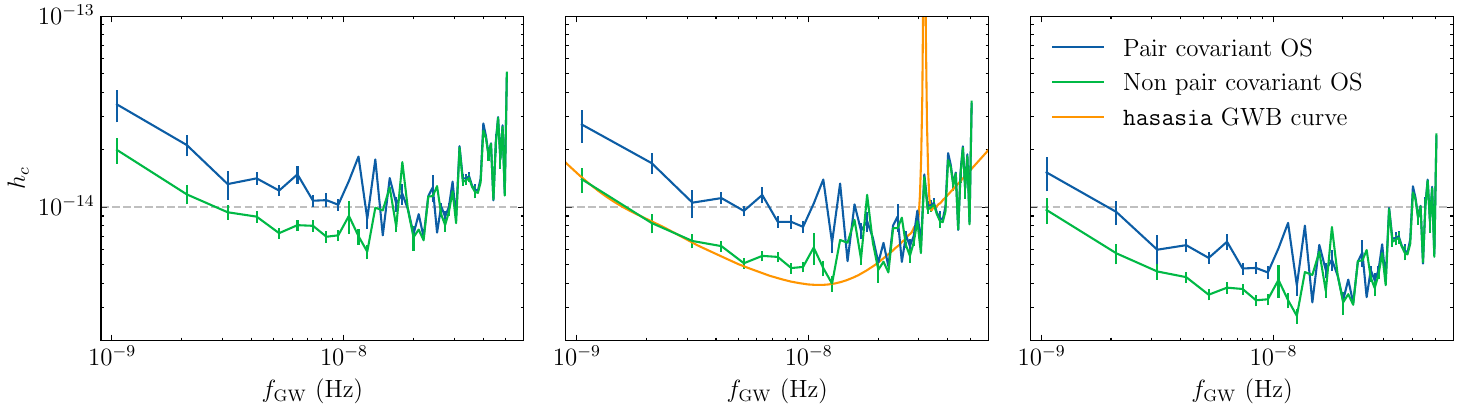}
    \caption{Pixel basis anisotropy sensitivity curves using a simulated dataset with 67 pulsars, 100 ns TOA uncertainty, and pulsars with baselines of identical length. These curves were generated by choosing a sky location for the left and right panels and averaging over all sky locations for the middle panel. The sky location of the left panel is a region of low pulsar density and correspondingly low sensitivity, whereas the sky location of the right panel is representative of the most sensitive sky location for the simulated PTA. This figure shows (i) validation against \texttt{hasasia} by comparing the map average to a \texttt{hasasia} GWB curve of the same dataset, (ii) the overestimated sensitivity when one neglects pair covariance, and (iii) the effect of sky location (or, more fundamentally, pulsar density) on the sensitivity of PTAs to anisotropy.}
    \label{fig:sensitivity_curves}
\end{figure*}

In Figure \ref{fig:sensitivity_curves}, we show sensitivity curves of a simulated PTA from the first set of simulations described in Section \ref{sec:simulations}. The map-averaged curves are consistent with the stochastic GWB sensitivity curves generated using \texttt{hasasia}. The curves that correspond to the most sensitive sky location of the PTAs are about a factor of 2-3 times more sensitive than the least sensitive sky location curves, which shows that the sky location of excess power in an anisotropic GWB can be an important factor in the time-to-detection.

Including the pulsar-pair covariance in the covariance matrix \citep{Allen23, pfos} has a significant impact on the sensitivity through an increase in the Fisher uncertainty for a given pixel, resulting in lower overall sensitivity.
The covariance of the correlations has to be accounted for because the null hypothesis in anisotropy searches is an isotropic GWB \citep{Gersbach25}, which induces covariant cross-correlations. Neglecting the covariance is appropriate in searches for an isotropic GWB however, as the null hypothesis in that case is noise. 
We find that at the lowest frequencies, the sensitivity is overestimated by a factor of two, and the overestimate reduces with higher frequencies until the ratio is approximately one. The agreement at higher frequencies is due to the 
magnitude of pair covariance scaling directly with the GWB power, which decreases at higher frequencies due to the assumed power-law behavior of the GWB.

\subsection{Scaling Laws} \label{sec:scaling}
The scaling law analyses were performed by computing the Fisher matrix in the pixel basis and spherical harmonic basis and grouping the results by the PTA parameter of interest, namely $N_{\mathrm{psr}}$, TOA uncertainty, and GW frequency, as well as angular scale for the spherical harmonics basis. We then fit a line to the log-log transformation of the square root of the diagonal elements of the Fisher matrix, $\sqrt{{\rm diag(}\mathbf{M})}$, versus the values of the parameter of interest. 
We choose to represent the scaling laws with respect to $\sqrt{{\rm diag(}\mathbf{M})}$ as it is easier to interpret than the corresponding uncertainties, with larger values representing higher sensitivity.
To condense all the sensitivity estimates to a single value for each parameter value, we take the median of the noise-marginalized distribution, i.e., the median of 1000 sensitivity estimates corresponding to 1000 noise draws, and then again take the median of the distribution across different realizations of the datasets. To fit a line to the data, we use \texttt{LMFIT} with asymmetric error bars on the data. The upper error bar is taken as the difference between the 75\textsuperscript{th} and 50\textsuperscript{th} percentiles, and the lower error bar is taken as the difference between the 50\textsuperscript{th} and 25\textsuperscript{th} percentiles.

\begin{figure*}
    \includegraphics[width=\textwidth]{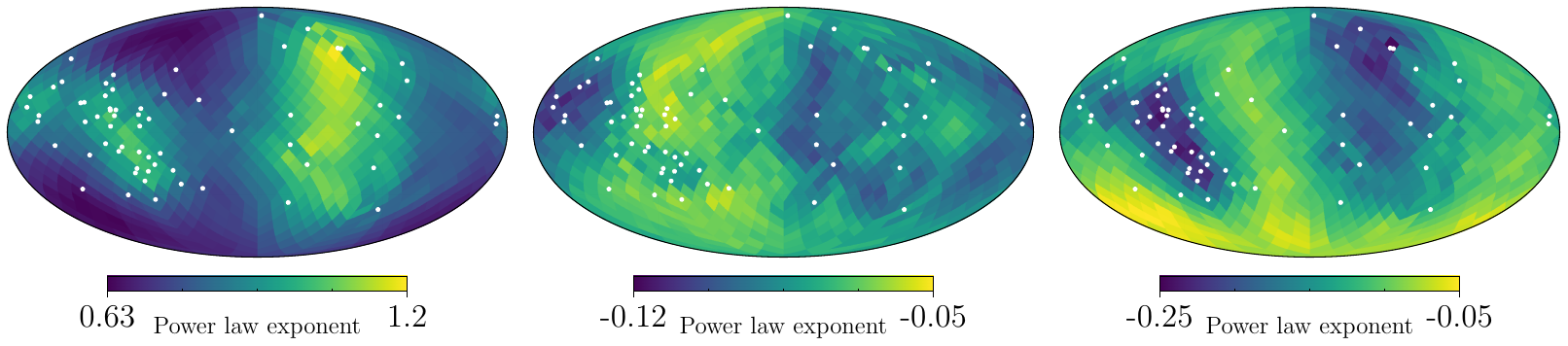}
    \caption{Pixel basis power law scaling exponents as a function of sky location and (\emph{left}) number of pulsars, (\emph{center}) TOA uncertainty, and (\emph{right}) GW frequency. The Fisher matrix diagonal elements increase with number of pulsars and decrease with TOA uncertainty and GW frequency.}
    \label{fig:scaling_pixels}
\end{figure*}

We show in Figure \ref{fig:scaling_pixels} the scaling of $\sqrt{{\rm diag(}\mathbf{M})}$ in the pixel basis with $N_{\mathrm{psr}}$, TOA uncertainty, and GW frequency. We find a median scaling law exponent of $\sqrt{{\rm diag(}\mathbf{M})}$ with $N_{\mathrm{psr}}$ of $0.8\pm0.1$, with the uncertainty being 1 standard deviation. The same analysis for TOA uncertainty and $f_\mathrm{GW}$ gives scaling law exponents of $-0.08\pm0.01$ and $-0.12\pm0.04$. We note that this dependence on $f_{\mathrm{GW}}$ is for the Fisher matrix only, 
and does not include the total contribution from the pulsar transmission functions \citep{hasasia}, encoded in the transformation to PSD or $h_c$. The total frequency dependence of the sensitivity in PSD space or $h_c$ space can be obtained by dividing this result by the frequency dependence of the PSD or equivalent characteristic strain in the PTA.

\begin{figure*}
    \includegraphics[width=\textwidth]{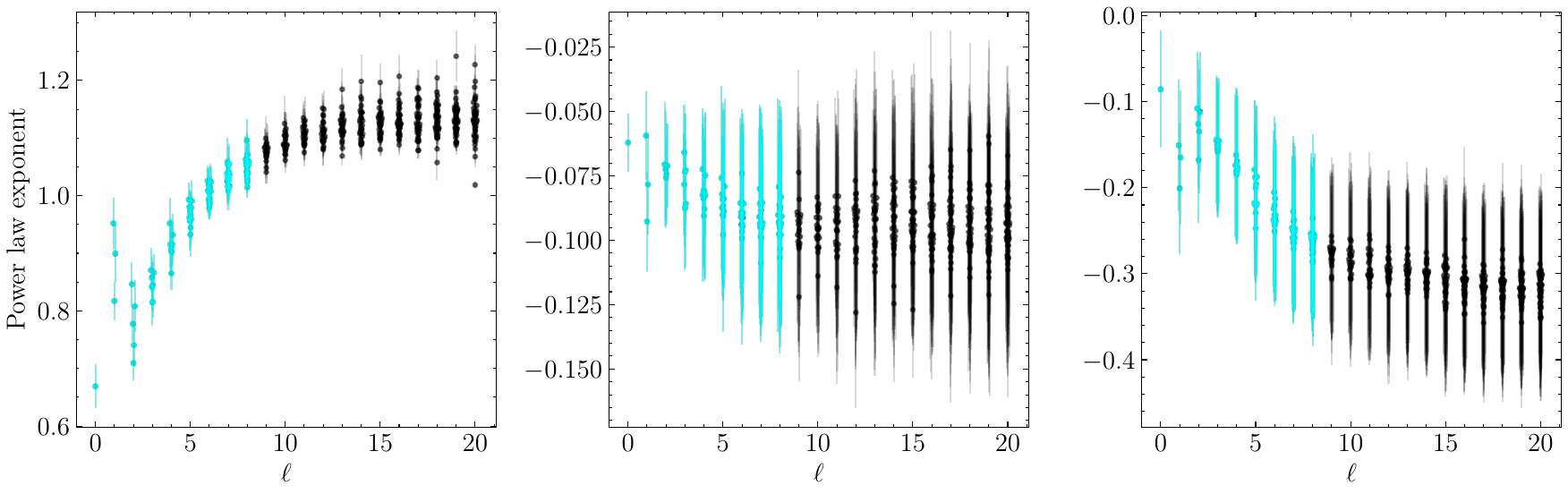}
    \caption{Spherical harmonic basis power law scaling exponents as a function of (\emph{left}) number of pulsars, (\emph{center}) TOA uncertainty, and (\emph{right}) GW frequency. The exponents are grouped by the $\ell$ value of the $c_{lm}$ coefficient, and the spread at each $\ell$ value corresponds to different $m$ values. $\sqrt{{\rm diag(}\mathbf{M})}$ scales with a positive exponent (i.e., increases) with number of pulsars and a negative exponent (i.e., decreases) with TOA uncertainty and GW frequency. The cyan markers correspond to an analysis performed setting $\ell_\mathrm{max}$ to 8. The black markers correspond to an analysis performed setting $\ell_\mathrm{max}$ to 20. The power law exponents generally take more extreme values at smaller angular scales.}
    \label{fig:scaling_clms}
\end{figure*}

\begin{figure*}
    \includegraphics[width=\textwidth]{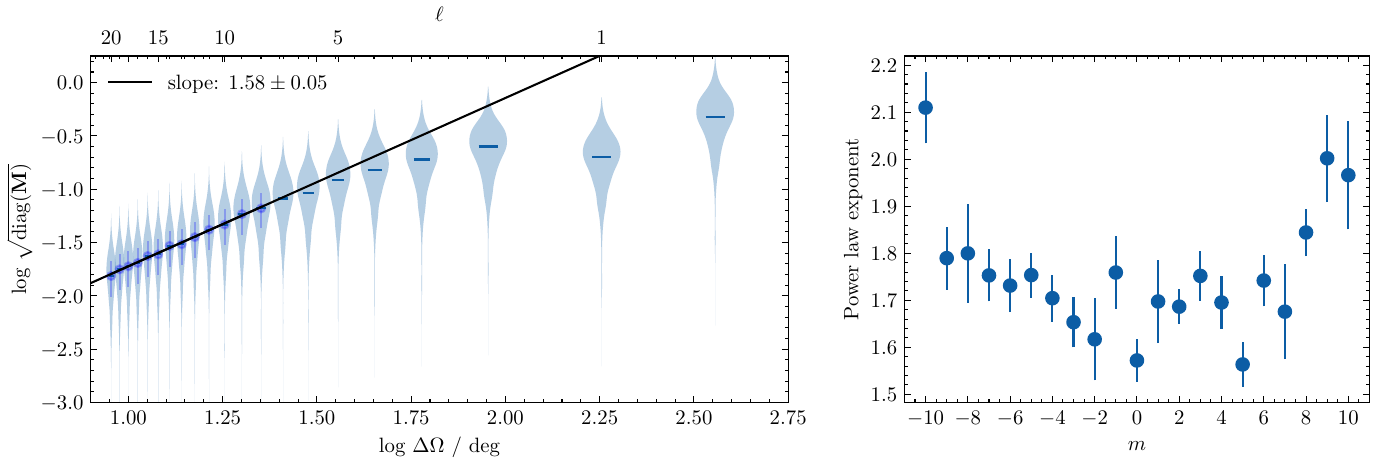}
    \caption{Sensitivity scaling in the spherical harmonic basis as a function of angular scale. Here we define angular scale as $180\degree$/$\ell$ for $\ell > 0$ and 360$\degree$ for $\ell=0$. The sensitivity increases with the angular scale of the anisotropy. \emph{Left.} The markers and error bars correspond to the medians and 25\textsuperscript{th}-75\textsuperscript{th} percentiles of the distributions. Only the Fisher matrix elements corresponding to coefficients with $8 \leq \ell \leq 20$ and $m=0$ were used in fitting the line. \emph{Right.} The power law exponents obtained by fixing $m$ to different values and considering only $10 \leq \ell \leq 20$. The dependence on $m$ is due to the anisotropic distribution of pulsars in the array and the fact that $m$ sets the orientation of the spherical harmonic. For example, $m=0$ corresponds to anisotropies at the north and south poles of the sphere (high and low declinations), while $m=10$ corresponds to anisotropies closer to the equator of the sphere for the range of $\ell$ values considered.}
    \label{fig:scaling_angular_scale}
\end{figure*}

In Figure \ref{fig:scaling_clms}, we show the spherical harmonic basis Fisher matrix scaling with $N_{\mathrm{psr}}$, TOA uncertainty, GW frequency. The scaling relations are similar to those in the pixel basis. In Figure \ref{fig:scaling_angular_scale}, we show the scaling with angular scale (given by $180\degree$/$\ell$ for $\ell > 0$ and 360$\degree$ for $\ell=0$). For this scaling, we use only the $c_{lm}$ coefficients with $8 \leq \ell \leq 20$ as lower values of $\ell$ follow a different, significantly shallower evolution. We verified that our results for this scaling law ($\sqrt{\mathrm{diag}(\mathbf{M})}\propto\Delta\Omega^\alpha$) do not change significantly whether holding $N_{\mathrm{psr}}$ fixed at 67 or including sensitivity results from all four $N_\mathrm{psr}$ values that were used in this work. The scaling law exponent for the selected range of $\ell$ values is $1.58\pm0.03$. The slope of the scaling law line does not change significantly when reducing the $\ell$ range to $10 \leq \ell \leq 20$ and including all values of $N_{\mathrm{psr}}$.

For ease of reference, we list our scaling relation results below.
\begin{align}
    \sqrt{\mathrm{diag}(\mathbf{M})}&\propto N_\mathrm{psr}^{0.8\pm0.1} \\
    \sqrt{\mathrm{diag}(\mathbf{M})}&\propto \delta t^{-0.08\pm0.01} \\
    \sqrt{\mathrm{diag}(\mathbf{M})}&\propto f_\mathrm{GW}^{-0.12\pm0.04}
\end{align}\label{eq:scaling}
The scaling with angular scale is more nuanced, so we refer the reader to Figure \ref{fig:scaling_angular_scale} and the related discussion in Section \ref{sec:discussion}.

\subsection{Forecasting PTA Sensitivity} \label{sec:forecast}
\begin{figure*}
    \includegraphics[width=\textwidth]{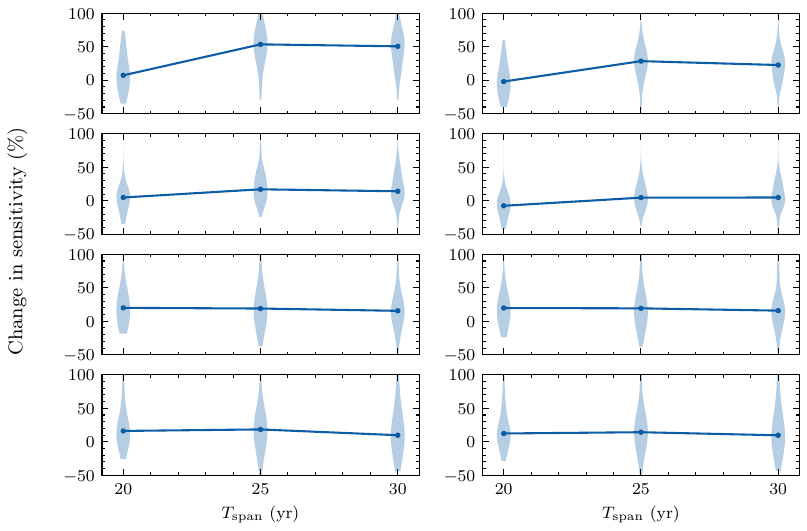}
    \caption{Percent change in sensitivity as a function of $T_\mathrm{span}$. All percentages are measured relative to the 16 yr baseline. The distributions contain the medians of each simulation, and the markers and lines represent the medians of the distributions. The rows represent different frequencies: $1/T_\mathrm{span}$ to $4/T_\mathrm{span}$ from top to bottom. The sky location is fixed to a pixel of relatively low (high) sensitivity in the left (right) column. Some changes in sensitivity extend beyond what is shown here, but we restrict the plot's limits to improve legibility. We note that these results include pulsar pair covariance. There is negligible dependence of sensitivity on $T_\mathrm{span}$ alone except at the lowest frequencies. This is because the $T_\mathrm{span}$ of individual pulsars grows with time and so pulsars which previously had too short of a baseline to contribute appreciable sensitivity to the lowest frequency bins begin to have greater impact on the sensitivity after being observed long enough to reach the lowest bins.}
    \label{fig:tspan_percent_change}
\end{figure*}

\begin{figure*}
    \includegraphics[width=\textwidth]{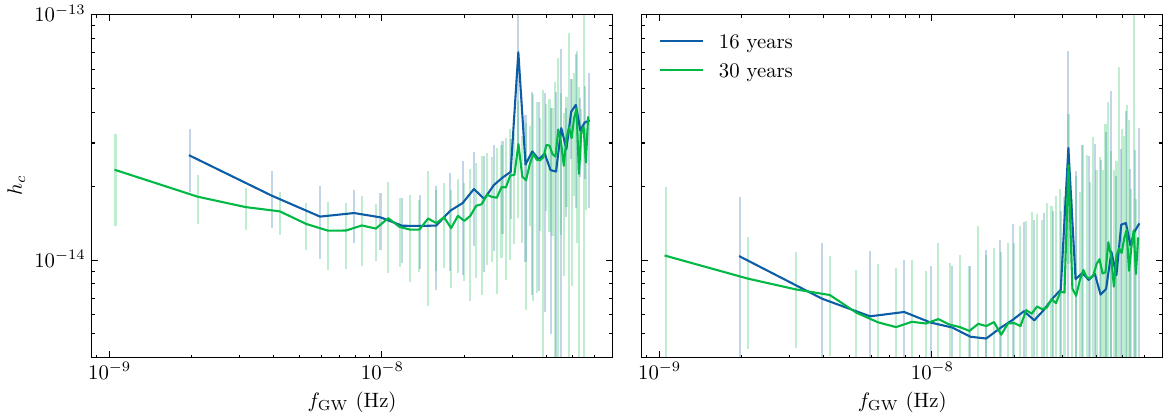}
    \caption{Sensitivity curves for simulated PTAs with realistic TOA uncertainty distributions and including pulsar pair covariance. The sky location was fixed to a pixel of relatively low (high) sensitivity in the left (right) panel. The curves represent the median sensitivity across datasets, and the error bars represent one standard deviation.}
    \label{fig:tspan_hc}
\end{figure*}

Using the realistic simulation set, we compute sensitivity curves for time-slices at 16, 20, 25, and 30 years. We compare the sensitivity at 16 years to the sensitivity at longer baselines as follows. 
First, we generate the sensitivity curves at the different time slices using the noise-marginalized, pair-covariant PFOS.
Next, we determine the sensitivity growth as a function of frequency. However, the frequency bins spaced at multiples of, e.g., $1/30\mathrm{yr}$ are not the same as those for $1/16\mathrm{yr}$. Therefore, to compare the sensitivity at each frequency, we select the multiple of $1/30\mathrm{yr}$ which is closest to each frequency bin of the 16-year simulations and compute the percent change in sensitivity relative to the 16-year sensitivity at each of the first 4 frequency bins. We then have a percent change in sensitivity for each $T_\mathrm{span}$, for each noise draw, for each frequency, and for each dataset. We show in Figure \ref{fig:tspan_percent_change} the noise-marginalized median changes in sensitivity as a function of $T_\mathrm{span}$. We also show the absolute sensitivities at 16 years and 30 years in Figure \ref{fig:tspan_hc} for a pixel with relatively low sensitivity and a pixel with relatively high sensitivity, but in this figure, we reduce the noise marginalization spread down to a point estimate (the median). The curves in Figure \ref{fig:tspan_hc} are then medians across realizations with error bars representing one standard deviation across realizations.

\section{Discussion} \label{sec:discussion}

From our sensitivity curves in Figure \ref{fig:sensitivity_curves}, we validate our method of making sensitivity curves against \texttt{hasasia} in the isotropic limit but now have a generalized method of quantifying PTA sensitivity to anisotropy which can be computed at any sky location or for any spherical harmonic basis function. Note that \citet{Mingarelli13, Taylor13, Mingarelli26} develop expressions for the expected angular power spectrum of a GWB, so one could use our method to quantify the sensitivity of a PTA to such power spectra. In this work, we also use our method to quantify the overestimate in sensitivity (roughly a factor of two at the lowest frequencies) when the covariance between pulsar-pair correlations is neglected. This overestimate is most significant at lower frequencies where the GWB signal is larger relative to the noise but insignificant at higher frequencies where the GWB is weak relative to the noise.

In addition, our simulated sensitivity curves show the large effect sky location has on PTA sensitivity to anisotropy, with the most sensitive pixel being a factor of $\sim2-3$ times more sensitive than the least sensitive pixel for the PTA configuration used. This effect is due to the anisotropic distribution of pulsar density in a PTA.

From our scaling law analyses for number of pulsars, TOA uncertainty, GW frequency, and angular scale, we found that the sensitivity of a PTA to anisotropy increases with the number of pulsars roughly linearly but with a strong dependence on pulsar density and angular scale. We also found that the sensitivity decreases with increasing TOA uncertainties but with a weaker dependence than the dependence on number of pulsars. Our results for number of pulsars and TOA uncertainty are in general agreement with the scaling laws found in \citet{Siemens13} for the optimal statistic (OS) signal-to-noise ratio (SNR); the relevant relations from that work which we discuss here is that the SNR scales as TOA uncertainty to the $-1/\gamma$ ($-0.23$ for $\gamma=13/3$) and scales linearly with the number of pulsars. Note that these scalings are valid in the intermediate-signal regime defined by \citet{Siemens13}, which is the regime in which our analysis was performed. The similarities between our results and those of \citet{Siemens13} are because the two detection methods are based on using the cross-correlations only of pulsar pairs. The anisotropy regression coefficients have a covariance matrix $(\mathbf{R}^T\mathbf{C}^{-1}\mathbf{R})^{-1}$. If we assume, for simplicity, a diagonal covariance matrix $\mathbf{C}$ of the cross-correlations with equal uncertainty for every cross-correlation, the anisotropy regression coefficients have uncertainties directly proportional to the uncertainties of the cross-correlations. This is the reason our result for the TOA uncertainty scaling is similar to the result from \citet{Siemens13}. However, the scaling we find is even weaker because our signal and null hypotheses are an anisotropic and isotropic GWB, respectively, whereas the signal and null hypotheses of the OS SNR are an isotropic GWB and noise, respectively. This is important because an isotropic GWB induces additional variance in the cross-correlations, and so our covariance matrix includes contributions from the GWB. This means part of the variance in the cross-correlations is unaffected by TOA uncertainty. In addition, we used only the lowest 6 bins of GW frequency for our scaling laws analyses, and the GWB is stronger at lower frequencies, so this likely resulted in even weaker scaling than if we had analyzed sensitivities at higher frequencies. To quantify approximately the contribution of pair covariance to the uncertainties in the anisotropic regression coefficients, we note that we found a factor of $\sim2$ difference in the sensitivity curves in Figure \ref{fig:sensitivity_curves}, so the GWB contributes a significant fraction of the total uncertainty.

As mentioned in the previous paragraph, \citet{Siemens13} found linear scaling of the OS SNR with $N_\mathrm{psr}$, which is in good agreement with our results in both bases. This strong dependence of sensitivity on number of pulsars is again due to the use of pulsar-pair cross-correlations for the frequentist anisotropy method employed here. \citet{Depta24} and \citet{Domcke25} found anisotropy sensitivity to scale as $\sqrt{N_\mathrm{psr}}$, in agreement with the isotropic GWB sensitivity scaling found by \citet{Babak24}, but this is because all of those works considered a different detection method which includes the auto-correlations. See \citet{AliHaimoud20} for additional discussion on anisotropy methods based on auto-correlations. \citet{Pol22} used a similar cross-correlation based method to our method in this work, although they studied the scaling of various SNRs and did not include pair covariance. They found linear scaling of total SNR (the maximum-likelihood ratio between an anisotropic GWB and noise) and isotropic SNR (isotropic GWB to noise) with $N_\mathrm{psr}$, which agrees with our result and that of \citet{Siemens13}. This is expected since these works all use cross-correlations only. \citet{Pol22} did not find significant scaling of anisotropic SNR with $N_\mathrm{psr}$, but this is because they injected an isotropic GWB, so, as they mentioned, the anisotropic SNR was not expected to scale.

We found that the sensitivity of a PTA to anisotropy generally decreases with decreasing angular scale. This is because there is a lower effective number of pulsars included in smaller angular scales. The dependence on angular scale is shallower at low values of $\ell$ and more extreme at high values of $\ell$. This is consistent with the results shown in Figure \ref{fig:scaling_clms} and is likely due to the finite number and anisotropic distribution of pulsars in the array, which cause the angular number density of the pulsar distribution relative to the spherical harmonic beams to change with $\ell$. If a PTA had a uniform distribution of many pulsars, the effective number of pulsars included in the characteristic area of a given spherical harmonic would scale the same as the area so that the density remains constant, giving linear scaling overall between area and effective number of pulsars. Since the angular scale of the spherical harmonic goes as $\ell^{-1}$, the area goes as $\ell^{-2}$, so linear scaling with area is equivalent to quadratic scaling with $\Delta\Omega$ and inverse quadratic scaling with $\ell$. The right panel of Figure \ref{fig:scaling_angular_scale} would then have markers all consistent with $(\Delta\Omega)^2$. However, as can be seen in that figure, there is also a dependence of the scaling on the values of $m$ under consideration, and this dependence is because $m$ sets the orientation of the spherical harmonic on the sphere. $m=0$ corresponds to the poles of the sphere, but there are few pulsars in the NANOGrav array at extreme declination. So, the scaling is shallower when $m$ is set to 0 compared to other values of $m$, and this effect can also be seen in Figure \ref{fig:scaling_pixels}, where areas of low pulsar density typically have shallower scaling than areas of high pulsar density.

Our results for $N_\mathrm{psr}$ and angular scale show the need for increasing the number of pulsars in PTA observing campaigns as well as increasing the sky coverage of PTAs for the largest improvements in sensitivity to anisotropy. Our results also show that combining PTA datasets through the International Pulsar Timing Array (IPTA, \citet{IPTA}) can greatly improve sensitivity to anisotropy as different PTAs have different sky coverage, so combining the datasets increases sky coverage besides increasing the number of pulsars. For example, doubling the number of pulsars through an IPTA combination would give roughly double sensitivity to anisotropy based on our scaling relation of sensitivity and number of pulsars. A future study of interest, given the stronger dependence of the sensitivity on number of pulsars than on noise, would be to quantify the gain in sensitivity PTAs could achieve by observing more pulsars but which have stronger noise levels.

For the other scaling law parameter we studied, GW frequency, we found that the diagonal elements of the Fisher matrix decrease with increasing GW frequency, although $h_c$ is also frequency-dependent. The total dependence of PTA sensitivity to anisotropy on GW frequency can be seen in Figure \ref{fig:sensitivity_curves}, and the frequency dependence is consistent with that of the sensitivity to an isotropic GWB using the methods of \citet{hasasia} despite the difference in methodology: \citet{hasasia} use pulsar auto-correlations rather than the cross-correlations that our work is based on. Our results also show that the net frequency dependence is approximately uniform across the sky as can be seen from the panels of Figure \ref{fig:sensitivity_curves}.

From our sensitivity forecast analysis in Section \ref{sec:forecast}, we found increasing only the observing timespan of a PTA from 16 years to 30 years has little effect on average on the sensitivity to anisotropy but with a large spread across realizations, sky location, and GW frequency. The dependence on sky location is physically only a dependence on pulsar density. The more sensitive sky location used in Figure \ref{fig:tspan_percent_change} (with a higher pulsar density) had, when using the mean across datasets of medians across noise draws, only a $\sim20\%$ growth in sensitivity at the lowest frequency bin by 30 years whereas the less sensitive sky location experienced a $\sim50\%$ growth for the same frequency bin by 30 years. To explain this, we first note that, as well as having lower pulsar density, the sky location with less sensitivity also had pulsars with shorter baselines nearby: only 3 pulsars had baselines beyond the second frequency bin ($\lesssim 4$ nHz) in that region of the sky in the 16 year timeslice but $\sim10$ pulsars by the 30 year timeslice. In contrast, the sky location with higher sensitivity had $\sim20$ pulsars nearby with baselines beyond 4 nHz in the 16 year timeslice and grew by 5-10 pulsars by the 30 year timeslice. Therefore, the higher growth in sensitivity in the less sensitive sky location compared to the more sensitive sky location is likely due to a larger relative increase in the ``effective'' number of pulsars which have an observing baseline long enough to contribute to the lowest frequencies.

Similar to the above argument but irrespective of sky location, we find that the sensitivity at the lower GW frequencies generally shows greater improvement than at higher frequencies, and the lowest frequency in particular shows the greatest improvement.
Across the whole PTA sky, the number of pulsars with a baseline beyond the second frequency bin is only 34 for the 16-year timeslice but all 67 pulsars by the 30-year timeslice, so this is likely introducing some of the scaling with $N_\mathrm{psr}$ into our results for $T_\mathrm{span}$ scaling.

Our results in Figures \ref{fig:tspan_percent_change} and \ref{fig:tspan_hc} show that increasing $T_\mathrm{span}$ without improving other parameters of a PTA has a much smaller effect on the sensitivity of the PTA to anisotropy than the effect of increasing $N_\mathrm{psr}$ (Section \ref{sec:scaling}) and the density of the pulsar distribution across the sky. Indeed, from Figure \ref{fig:tspan_hc} it can be seen that a PTA with 16 years of observation is up to twice as sensitive to anisotropy in sky locations with high pulsar density than a PTA with 30 years of observation is to anisotropy in sky locations with low pulsar density. This suggests a large gain in sensitivity to anisotropy that PTAs can achieve with greater pulsar density.

As a final note on our sensitivity forecasts, the increases in sensitivity reported here are conservative as we maintained the number and sky coverage of pulsars, the number of telescopes, and the TOA uncertainties that were present in the NANOGrav 15-year dataset.

\section{A New Multi-Resolution Pixel Basis}\label{sec:mesh}

Motivated by the strong dependence of PTA sensitivity on sky location (as shown in Figure \ref{fig:sensitivity_curves}), we introduce a multi-resolution pixel basis. We use the multi-resolution HEALPix functionality \citep{Singer16, mhealpyPaper} in the software package \texttt{mhealpy} \citep{mhealpySoftware} and adapt it to PTA anisotropy methods. In brief, the multi-resolution HEALPix tessellation works by assigning a unique index to every possible pixel. The index encodes the pixel's location and resolution. This is called the UNIQ scheme, and it works similarly to the NESTED scheme, but the NESTED and RING schemes encode the resolution using the total number of pixels in the map. This is not possible with a multi-resolution tessellation, so the UNIQ scheme removes that implicit dependence on the number of pixels. Explicitly, the index of a pixel in the UNIQ scheme is related to the index of the equivalent pixel of a map of resolution $N_{\mathrm{side}}$ in the NESTED scheme as $i_{\mathrm{UNIQ}} = i_{\mathrm{NESTED}} + 4N_{\mathrm{side}}^2$.
See \citet{Fernique14, Reinecke15, Youngren17} for more details on multi-resolution HEALPix tessellations.

PTA anisotropy searches with the multi-resolution pixel basis work the same way as searches with the single-resolution pixel basis except we use a general normalization ($\Delta\Omega_k$/$4\pi$) rather than $N_{\mathrm{pix}}^{-1}$. Using $N_{\mathrm{pix}}^{-1}$ assumes all pixels have the same area, so that normalization is valid for the single-resolution pixel basis but not for the multi-resolution pixel basis. Explicitly, the response matrix is
\begin{equation}
    R_{ab}^k = \frac{\Delta\Omega_k}{4\pi}\mathcal{R}_{ab}(\hat\Omega_k),
\end{equation}
and the form of the maximum likelihood estimator is unchanged.

To verify the functionality of the new basis, we perform injection-and-recovery simulations with isotropic and anisotropic GWB injections using analytical recoveries. However, we restrict the analyses to small numbers of pixels because we find that, when the number of pixels is large, the Fisher matrix often becomes nearly singular, so the analytical solution $\mathbf{P}=\mathbf{M}^{-1}\mathbf{X}$ becomes numerically unstable due to the need to invert the Fisher matrix. This instability is mostly due to strong covariances between pixels, but other effects contribute as well, such as anisotropic pulsar distribution and covariant cross-correlations. For a simplistic simulation with 67 uniformly distributed simulated pulsars, the Fisher matrix for the pixel basis at $N_\mathrm{side}=8$ has a condition number $\mathcal{O}(10^{14})$, but setting the off-diagonal elements to zero (which defines the \emph{radiometer} pixel basis, the basis used in the previous sections of this work) reduces the condition number to $\mathcal{O}(1)$. This suggests the covariances between pixels causes most of the instability. Additionally, since the cross-correlations are not independent, the number of pixels that can be constrained is less than a simple counting argument limit like $N_\mathrm{pix}\lesssim N_\mathrm{cc}$. We also find that by using the NANOGrav distribution of pulsar locations, the condition number increases from $\mathcal{O}(10^{14})$ to $\mathcal{O}(10^{18})$, which indicates that sky coverage contributes significantly. Numerical instabilities can be overcome using standard methods of dealing with nearly singular matrices, e.g., SVD regularization, using a forward-modeled approach to avoid the ill-posed inverse problem, etc. However, we choose to simply use few pixels and side-step the problem entirely as our aim here is only to demonstrate the new basis rather than actually perform an anisotropy search.

\begin{figure}
    \includegraphics[width=\columnwidth]{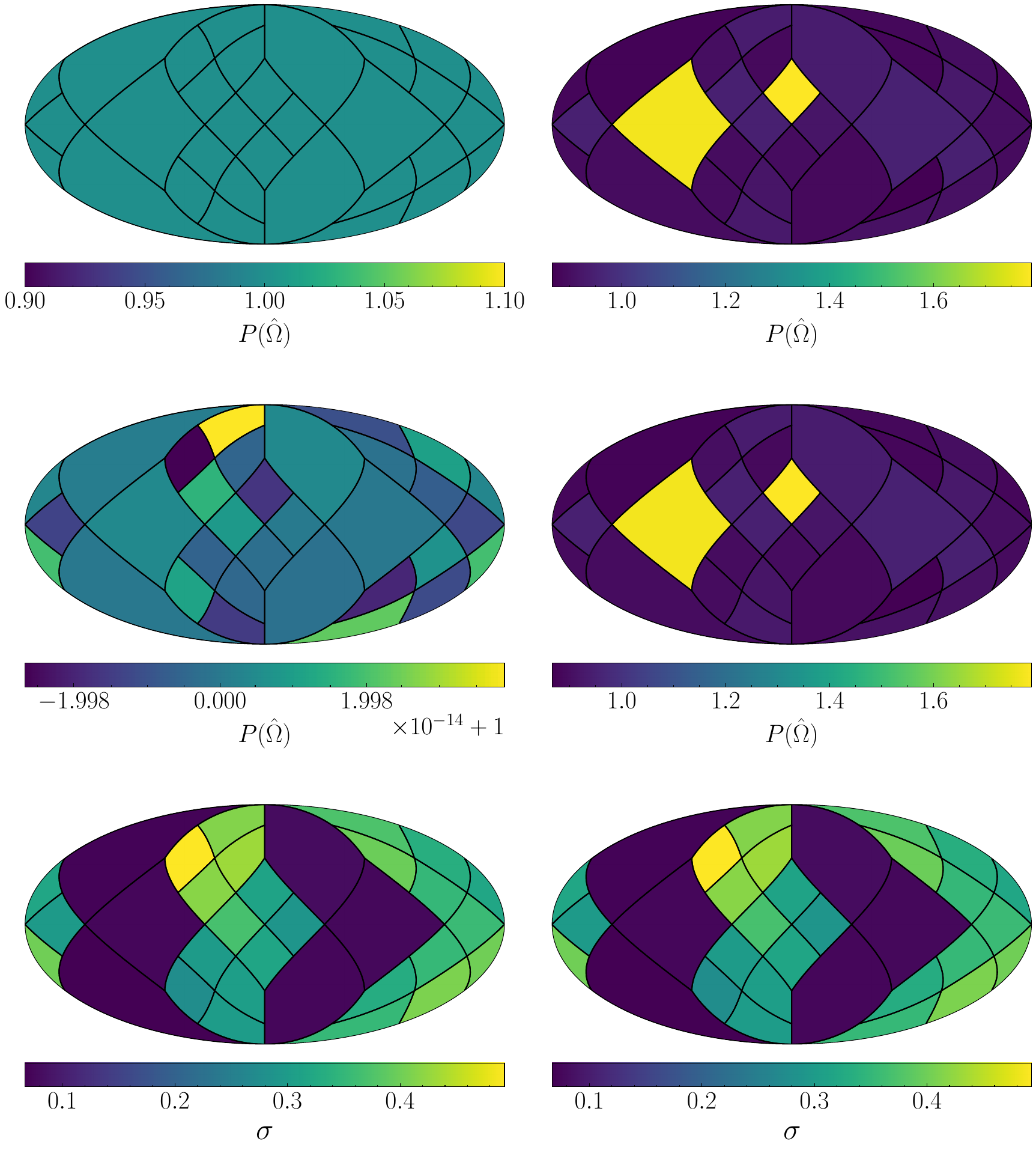}
    \caption{Analytical recoveries of simulated injected GWB power distributions. \textit{Top.} The injected GWB power distributions with an isotropic GWB in the left column and an anisotropic GWB in the right column. \textit{Middle.} The recovered sky maps with the multi-resolution pixel basis. \textit{Bottom.} Uncertainties in the recoveries, estimated as the square root of the diagonal elements of the inverse Fisher matrix. The uncertainties are the same between the left and right columns of the figure because the Fisher matrix is independent of the injected GWB since we neglected pair covariance in these multi-resolution simulations. The large difference in uncertainties between pixels of different areas is because we are solving the maximum likelihood problem for power density $P$. If we instead solve for a relative amplitude $P\Delta\Omega$, the uncertainties would be approximately uniform.}
    \label{fig:moc_analytical}
\end{figure}

To simulate cross-correlations, we generate 75 pulsars randomly distributed across the sky, and compute the cross-correlations induced by the injected GWB by multiplying the simulated PTA response matrix with the injected power vector. We use cross-correlation uncertainties of 0.1 for all pulsar pairs (for reference, the HD curve has a maximum value of 0.5), and we generate the recovered sky maps using the maximum likelihood estimator $\mathbf{P}=\mathbf{M}^{-1}\mathbf{X}$. We perform tests with an isotropic injection and an anisotropic injection. The isotropic GWB was simulated to have power $P_k=1 \ \forall k$ where $k$ represents the pixel index, while the anisotropic GWB was generated by starting with an isotropic GWB, doubling the power in two pixels, adding a random number between 0 and 0.1 to each pixel, and normalizing the result. We show the injections, recoveries, and uncertainties in Figure \ref{fig:moc_analytical}. The difference between injection and recovery are $\mathcal{O}(10^{-14})$, much smaller than the $\mathcal{O}(1)$ GWB, indicating successful recoveries.

\begin{figure*}
    \includegraphics[width=\textwidth]{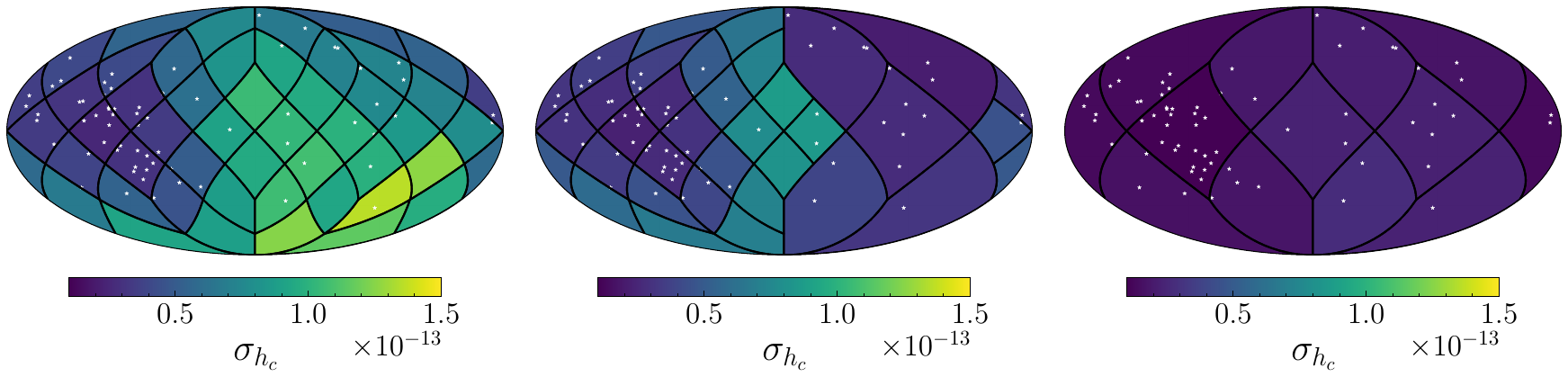}
    \caption{Uncertainty maps for (\textit{left}) a single-resolution $N_\mathrm{side}=2$ parameterization, (\textit{center}) a multi-resolution parameterization, and (\textit{right}) a single-resolution $N_\mathrm{side}=1$ parameterization. The three colorbars have the same limits for easier comparison.}
    \label{fig:moc_sensitivity_maps}
\end{figure*}

In Figure \ref{fig:moc_sensitivity_maps}, we show a sensitivity map in this basis compared to sensitivity maps produced in the single-resolution pixel basis. These sensitivity maps were produced using one of the simulated datasets from the scaling law analysis. The $N_\mathrm{side}=2$ map in the left panel has higher uncertainties but better resolution while the $N_\mathrm{side}=1$ map in the right panel has lower uncertainties but also lower resolution. The multi-resolution basis allows us to fuse the two resolutions and corresponding uncertainties to balance the uncertainties across the sky. Using this new basis, we can effectively trade localization for uncertainty and vice versa at each region of the sky independently of other regions of the sky. This allows us to take advantage of the anisotropic distribution of pulsars in PTAs to achieve better resolution. This also highlights the importance of increasing the number of pulsars in PTAs irrespective of sky coverage as we can increase the resolution wherever the new pulsars are added. Note, however, only the uncertainties on power density are affected by using the multi-resolution basis. As mentioned in the caption of Figure \ref{fig:moc_analytical}, the uncertainties on pixel amplitudes cannot be tuned in the same way.

Although we have introduced and verified the multi-resolution basis in this section and produced sensitivity maps, this basis still needs more development before it is complete. In particular, the statistic used to produce a multi-resolution mesh for a given dataset needs to be explored, and the basis needs to be tested with more realistic simulations starting from the TOA level rather than from the pulsar-pair cross-correlations as was done in this work. This basis was motivated by the anisotropic sensitivity of PTAs and allows for the exchange of localization for uncertainty (and vice versa) in each location of the sky, so a natural statistic used to set resolutions is the sensitivity of the PTA as we demonstrated in Figure \ref{fig:moc_sensitivity_maps}; in future work, however, we plan to use reversible-jump MCMC (RJMCMC; \citet{Green95}) methods to select a preferred mesh for a given PTA dataset. The Fisher matrix used in this work only accounts for noise and geometry (through $\mathbf{C}$ and $\mathbf{R}$, respectively), so more data-driven methods like RJMCMC would be desirable to include contributions from the likelihood as well. A question to explore afterwards is whether the multi-resolution basis can mitigate bias from angular scales of the GWB smaller than the cutoff resolution used for the reconstruction. \citet{Semenzato25} argued that, independent of basis, anisotropies at angular scales beyond the reconstruction's cutoff resolution bias the reconstruction. \citet{Agarwal26} showed that, in the spherical harmonic basis, making  $\ell_\mathrm{max}$ sufficiently large avoids this bias, and they use a method similar to the local resolution spherical harmonic method of \citet{Grunthal26}. The multi-resolution pixel basis can be used to achieve an analogous local resolution method to \citet{Grunthal26} but in the pixel basis, so it may likewise have the additional advantage of mitigating bias from small angular scales. This is beyond the scope of this work, however, and further research would be required to make a definitive statement.

\begin{acknowledgments}
T.T.M. thanks Kushagra N. Nag and Zach Zelensky for helpful discussions. T.T.M. is grateful for support from the Texas Tech University Department of Physics \& Astronomy through a Departmental Research Associate Fellowship. N.S.P. acknowledges support from startup funds from Texas Tech University. The authors acknowledge the High Performance Computing Center (HPCC) at Texas Tech University for providing computational resources that have contributed to the research results reported within this paper. Some of the results in this paper have been derived using the \texttt{healpy} and \texttt{HEALPix} packages.
\input{acknowledgments.tex}
\end{acknowledgments}

\begin{contribution}
T.T.M. performed the analyses, produced the plots, and wrote the manuscript. K.A.G. implemented the per-frequency optimal statistic and pair covariance frameworks into \texttt{MAPS}. N.S.P. conceived of the research idea and provided guidance throughout the project. N.S.P. also performed the noise analysis to get the values listed in Appendix \ref{appendix:injected_noise}. C.M.F.M. provided insights in interpreting some of the results.
\end{contribution}

\software{
          \texttt{Defiant} \citep{pfos},
          \texttt{ENTERPRISE} \citep{enterprise},
          \texttt{enterprise\_extensions} \citep{enterprise_extensions},
          \texttt{h5py} (h5py.org),
          \texttt{hasasia} \citep{hasasia_software},
          \texttt{healpy} \citep{Zonca2019},
          \texttt{JAX} \citep{jax2018github},
          \texttt{la\_forge} \citep{laforge},
          \texttt{libstempo} (https://github.com/vallis/libstempo),
          \texttt{LMFIT} \citep{LMFIT},
          \texttt{MAPS} \citep{Pol22},
          \texttt{Matplotlib} \citep{Hunter:2007},
          \texttt{mhealpy} \citep{mhealpySoftware},
          \texttt{NumPy} \citep{harris2020array},
          \texttt{PTMCMCSampler} \citep{ptmcmcsampler},
          \texttt{SciPy} \citep{SciPy}
}

\bibliography{sensitivity}{}
\bibliographystyle{aasjournalv7}

\appendix
\section{Injected Noise Parameters}
\label{appendix:injected_noise}
\centering
\begin{table}[h!]
\begin{tabular} {|c|c|c||c|c|c|}
    \hline
    Pulsar & Amplitude & Spectral Index & Pulsar & Amplitude & Spectral Index \\
    \hline
    B1855+09 & -13.900 & 3.613 & J1730-2304 & -12.974 & $5\times10^{-4}$ \\
    B1937+21 & -13.530 & 3.853 & J1738+0333 & -14.860 & 5.524 \\
    B1953+29 & -12.740 & 1.632 & J1741+1351 & -13.637 & 1.414 \\
    J0023+0923 & -13.320 & 0.091 & J1744-1134 & -13.612 & 0.469 \\
    J0030+0451 & -17.740 & 0.918 & J1745+1017 & -11.865 & 2.321 \\
    J0340+4130 & -15.875 & 6.994 & J1747-4036 & -12.565 & 2.483 \\
    J0406+3039 & -14.573 & 6.915 & J1751-2857 & -18.778 & 4.342 \\
    J0437-4715 & -13.974 & 0.465 & J1802-2124 & -12.209 & 0.921 \\
    J0509+0856 & -12.117 & $1 \times 10^{-4}$ & J1811-2405 & -19.184 & 4.743 \\
    J0557+1551 & -14.768 & 6.977 & J1832-0836 & -13.765 & 2.019 \\
    J0605+3757 & -15.810 & 4.304 & J1843-1113 & -13.325 & 0.011 \\
    J0610-2100 & -12.708 & 2.907 & J1853+1303 & -13.320 & 0.881 \\
    J0613-0200 & -13.455 & 1.742 & J1903+0327 & -12.201 & 1.585 \\
    J0636+5128 & -13.357 & 0.025 & J1909-3744 & -14.029 & 0.003 \\
    J0645+5158 & -13.248 & 0.502 & J1910+1256 & -13.547 & 2.266 \\
    J0709+0458 & -14.162 & 6.993 & J1911+1347 & -17.523 & 1.227 \\
    J0740+6620 & -13.589 & 0.010 & J1918-0642 & -13.534 & 2.050 \\
    J0931-1902 & -16.794 & 2.424 & J1923+2515 & -16.066 & 6.996 \\
    J1012+5307 & -12.629 & 0.555 & J1944+0907 & -13.245 & 1.971 \\
    J1012-4235 & -12.445 & 2.812 & J1946+3417 & -12.475 & 0.852 \\
    J1022+1001 & -12.107 & 0.001 & J2010-1323 & -16.297 & 6.993 \\
    J1024-0719 & -13.226 & 0.188 & J2017+0603 & -19.939 & 0.550 \\
    J1125+7819 & -12.843 & 0.008 & J2033+1734 & -15.626 & 6.998 \\
    J1312+0051 & -18.462 & 4.150 & J2043+1711 & -13.972 & 0.700 \\
    J1453+1902 & -18.426 & 4.801 & J2124-3358 & -15.093 & 6.984 \\
    J1455-3330 & -13.261 & 1.789 & J2145-0750 & -12.884 & 0.404 \\
    J1600-3053 & -13.496 & 1.378 & J2214+3000 & -12.924 & 0.008 \\
    J1614-2230 & -16.108 & 6.994 & J2229+2643 & -15.841 & 6.996 \\
    J1630+3734 & -14.238 & 6.982 & J2234+0611 & -13.538 & 0.019 \\
    J1640+2224 & -16.642 & 6.983 & J2234+0944 & -15.954 & 6.997 \\
    J1643-1224 & -12.267 & 0.489 & J2302+4442 & -15.730 & 6.990 \\
    J1705-1903 & -12.111 & 0.002 & J2317+1439 & -13.532 & 0.246 \\
    J1713+0747 & -14.085 & 2.515 & J2322+2057 & -19.890 & 3.228 \\
    J1719-1438 & -18.869 & 3.209 & All (GWB) & -14.620 & 4.333 \\
    \hline
\end{tabular}
\caption{The spectral parameters of the injected power-law red noise in the simulated datasets.}
\label{table:injections}
\end{table}
\end{document}

%% file: authors.tex
\author[0000-0001-5134-3925]{Gabriella Agazie}
\affiliation{Center for Gravitation, Cosmology and Astrophysics, Department of Physics and Astronomy, University of Wisconsin-Milwaukee,\\ P.O. Box 413, Milwaukee, WI 53201, USA}
\email{}
\author[0000-0001-7544-7876]{Nikita Agarwal}
\affiliation{Department of Physics and Astronomy, West Virginia University, P.O. Box 6315, Morgantown, WV 26506, USA}
\affiliation{Center for Gravitational Waves and Cosmology, West Virginia University, Chestnut Ridge Research Building, Morgantown, WV 26505, USA}
\email{}
\author[0000-0002-8935-9882]{Akash Anumarlapudi}
\affiliation{Department of Physics and Astronomy, University of North Carolina, Chapel Hill, NC 27599, USA}
\email{}
\author[0000-0003-0638-3340]{Anne M. Archibald}
\affiliation{Newcastle University, NE1 7RU, UK}
\email{}
\author[0009-0008-6187-8753]{Zaven Arzoumanian}
\affiliation{X-Ray Astrophysics Laboratory, NASA Goddard Space Flight Center, Code 662, Greenbelt, MD 20771, USA}
\email{}
\author[0000-0002-8395-957X]{Anjana Ashok}
\affiliation{Department of Physics, Oregon State University, Corvallis, OR 97331, USA}
\email{}
\author[0000-0002-4972-1525]{Jeremy G. Baier}
\affiliation{Department of Physics, Oregon State University, Corvallis, OR 97331, USA}
\email{}
\author[0000-0003-2745-753X]{Paul T. Baker}
\affiliation{Department of Physics and Astronomy, Widener University, One University Place, Chester, PA 19013, USA}
\email{}
\author[0000-0003-0909-5563]{Bence B\'{e}csy}
\affiliation{Institute for Gravitational Wave Astronomy and School of Physics and Astronomy, University of Birmingham, Edgbaston, Birmingham B15 2TT, UK}
\email{}
\author[0000-0002-2183-1087]{Laura Blecha}
\affiliation{Physics Department, University of Florida, Gainesville, FL 32611, USA}
\email{}
% ...

%% file: acknowledgments.tex
The work of A.As., R.B., R.C., X.S., J.T., and D.W.\ is partly supported by the George and Hannah Bolinger Memorial Fund in the College of Science at Oregon State University.
A.As.\ gratefully acknowledges the support of Moore Foundation.
L.B.\ acknowledges support from the National Science Foundation under award AST-2307171 and from the National Aeronautics and Space Administration under award 80NSSC22K0808.
P.R.B.\ is supported by the Science and Technology Facilities Council, grant number ST/W000946/1.
S.B.\ gratefully acknowledges the support of a Sloan Fellowship, and the support of NSF under award \#1815664.
M.C.\ acknowledges support by the European Union (ERC, MMMonsters, 101117624).
Support for this work was provided by the NSF through the Grote Reber Fellowship Program administered by Associated Universities, Inc./National Radio Astronomy Observatory.
Pulsar research at UBC is supported by an NSERC Discovery Grant and by CIFAR.
K.C.\ and M.S.T.\ are supported by a UBC Four Year Fellowship.
T.D.\ and M.T.L.\ received support by an NSF Astronomy and Astrophysics Grant (AAG) award number 2009468 during this work.
E.C.F.\ is supported by NASA under award number 80GSFC24M0006.
K.A.G.\ and S.R.T.\ acknowledge support from an NSF CAREER award \#2146016.
D.C.G.\ is supported by NSF Astronomy and Astrophysics Grant (AAG) award \#2406919.
A.D.J.\ acknowledges support from the Caltech and Jet Propulsion Laboratory President's and Director's Research and Development Fund.
A.D.J.\ acknowledges support from the Sloan Foundation.
N.La.\ was supported by the Vanderbilt Initiative in Data Intensive Astrophysics (VIDA) Fellowship.
Part of this research was carried out at the Jet Propulsion Laboratory, California Institute of Technology, under a contract with the National Aeronautics and Space Administration (80NM0018D0004).
D.R.L.\ and M.A.M.\ are supported by NSF \#1458952.
M.A.M.\ is supported by NSF \#2009425.
C.M.F.M.\ was supported in part by the National Science Foundation under Grants No.\ NSF PHY-1748958 and NASA LPS 80NSSC24K0440. C.M.F.M.\ also thanks the Center for Computational Astrophysics (CCA) of the Flatiron Institute for support. The Flatiron Institute is supported by the Simons Foundation.
A.Mi.\ acknowledges support from a Royal Society University Research Fellowship (URF-R1-251896)
The Dunlap Institute is funded by an endowment established by the David Dunlap family and the University of Toronto.
K.D.O.\ was supported in part by NSF Grant No.\ 2207267.
T.T.P.\ acknowledges support from the Extragalactic Astrophysics Research Group at E\"{o}tv\"{o}s Lor\'{a}nd University, funded by the E\"{o}tv\"{o}s Lor\'{a}nd Research Network (ELKH), which was used during the development of this research.
P.P.\ and S.R.T.\ acknowledge support from NSF AST-2007993.
H.A.R.\ is supported by NSF Partnerships for Research and Education in Physics (PREP) award No.\ 2216793.
S.M.R.\ and I.H.S.\ are CIFAR Fellows.
Portions of this work performed at NRL were supported by ONR 6.1 basic research funding.
J.D.R.\ also acknowledges support from start-up funds from Texas Tech University.
S.C.S.\ and S.J.V.\ are supported by NSF award PHY-2011772.
J.S.\ is supported by an NSF Astronomy and Astrophysics Postdoctoral Fellowship under award AST-2202388, and acknowledges previous support by the NSF under award 1847938.
J.P.W.V.\ acknowledges support from NSF AccelNet award No.~2114721.
O.Y.\ is supported by the National Science Foundation Graduate Research Fellowship under Grant No.\ DGE-2139292.